# Impact of the topology of global macroeconomic network on the spreading of economic crises

Kyu-Min Lee,[1] Jae-Suk Yang,[2] Gunn Kim,[3] Jaesung Lee,[4] Kwang-Il Goh,[1,*] and In-mook Kim[1]

[1]Department of Physics, Korea University, Seoul 136-713, Korea
[2]The Sanford C. Bernstein & Co. Center for Leadership and Ethics, Columbia Business School, Columbia University, New York, NY 10027, USA
[3]Department of Physics, Sejong University, Seoul 143-747, Korea
[4]Department of Mathematics, Sogang University, Seoul 121-742, Korea
*Corresponding author. E-mail: kgoh@korea.ac.kr

Throughout economic history, the global economy has experienced recurring crises. The persistent recurrence of such economic crises calls for an understanding of their generic features rather than treating them as singular events. Global economic system is a highly complex system and can best be viewed in terms of a network of interacting macroeconomic agents. In this regard, from the perspective of collective network dynamics, here we explore how the topology of global macroeconomic network affects the patterns of spreading of economic crises. Using a simple toy model of crisis spreading, we demonstrate that an individual country's role in crisis spreading is not only dependent on its gross macroeconomic capacities, but also on its local and global connectivity profile in the context of the world economic network. We find that on one hand clustering of weak links at the regional scale can significantly aggravate the spread of crises, but on the other hand the current network structure at the global scale harbors higher tolerance of extreme crises compared to more "globalized" random networks. These results suggest that there can be a potential hidden cost in the ongoing globalization movement towards establishing less-constrained, trans-regional economic links between countries, by increasing vulnerability of the global economic system to extreme crises.

**Introduction**

Historic events, from the Great Depression and the collapse of the Bretton Woods system to the more recent Asian financial crisis and the current economic turmoil triggered by the United States housing market, are only some of the dramatic episodes in the chronicle of global economic crises [1]. Economic crises are diverse in scale and detail, and recur intermittently yet persistently. Researchers in political economy and international relations have devoted this problem for many years [2]. The persistent recurrence of economic crises despite great economic and political efforts to prevent them suggests that they might be a generic rather than singular part of the global economic dynamics [3, 4], and there might be a general pattern beyond episode-specific particulars. Underlying this rationale is the idea that the global economic system is a highly interlinked and interactive many-body complex system comprised of heterogeneous entities (countries) with intricate interdependencies [5]. Along this line, here we address this problem in a quantitative setting using tools and concepts from network science and viewing the global economic crisis from the perspective of collective network dynamics [6-8].

From the spreading of infectious diseases [9] to political opinions [10], the substrate structure over which pathogens or opinions are transmitted plays a crucial role in determining the large-scale pattern of spreading dynamics [8, 11]. In the global economic system, countries and other economic institutions are connected through various economic channels ranging from commodity and capital trading to political and administrative partnerships. Changes in a country's economic conditions can spread through these connection channels in diverse ways, either by directly inducing supply/demand imbalance and liquidity risk or by indirectly affecting the comparative advantage and mutual credit landscapes [12]. Accordingly, connectivity patterns of the global economic system can impose structural constraints on the dynamics of crisis spreading.

Network properties of the global economic system have been documented in many studies, which have revealed scale-free weighted topology at various levels [13-16], increasing connectivity over time [17-20], and implications for economic development [21, 22]. However, much less is known about how such structural properties could translate into or form constraints on dynamic processes occurring in the global economic system, especially with respect to crisis spreading. In this study we attempt to fill this gap by establishing a general picture upon which more specific and context-dependent situations can be studied and understood.

## Materials and Methods

The global macroeconomic network

Different connection channels in the global economic system, such as commodity trading and administrative partnerships, may possess distinct structural profiles and dynamic response characteristics depending on various economic and political constraints. To model crisis spreading in complete detail, every economic channel should be incorporated individually. It is, however, neither possible nor necessary to do that in our study because our main goal is to build a general picture and illustrate key features that are less dependent on microscopic details, using the philosophy of critical phenomena in statistical physics. Different economic channels, despite their diversity, are not completely independent of one another, as they are under a common economic and political environment to some extent [23]. It has long been argued that among the diverse channels the trading relationships represent the most direct and consequential economic interdependency between countries [24, 25]. From a practical viewpoint, international trading has been most comprehensively and transparently documented by official agencies for many years. Therefore, we used international trade data to build the global macroeconomic network (GMN) in this study.

The GDP data was obtained from the International Monetary Fund World Economic Outlook Database October 2008 edition (http://www.imf.org/external/pubs/ft/weo/2008/02/weodata/index.aspx). From the database, we used the five-year list of each country's GDP, from the year 2002 to 2006, except for 13 countries with limited data for various reasons. Whenever possible, we used a five-year average value for the country's current representative GDP value (in US dollars). Otherwise, we used the average value for the available period.

The trade data was obtained from the United Nations International Merchandise Trade Statistics Publication Yearbook 2007 (http://comtrade.un.org/pb/CountryPages.aspx?y=2007). This dataset contains for each country, among others, a list of trading partners and corresponding trade volumes, both for exports and imports, in the same five-year period from 2002 to 2006. As with the GDP, we used five-year average value for the representative trade volume between two countries. Otherwise, we used the average value for the available period.

In order to build GMN, we considered only countries with both the GDP and trade data available in our datasets, resulting in 175 countries. The list of 175 countries with their characteristics relevant to this study is tabulated in Table S1. In this way, we have the GMN as a directed, node- and link-weighted network of 175 countries.

Crisis spreading model

From a theoretical perspective, the spread of an economic crisis can be framed as a cascading failure or avalanche process [26-30]. Along this line, we introduce a toy model of crisis spreading on GMN as follows (Figs. 1 and S1). Each node has its capacity $C_i$ and each link has the weight $W_{ij}$ (Fig. 1A). Suppose that a country $i$ collapses (Fig. 1B). As a result, the weights $W$ of all links of country $i$ are decreased by a fraction $f$. If the total decrement $\Delta W$ of either the incoming or outgoing link weights of any countries connected to the collapsed country exceeds a fraction $t$ of its node capacity $C$, then these countries also collapse (for example, country $k$ in Fig. 1C), thereby decreasing their link weights by fraction $f$ and initiating an avalanche of collapses. This in turn can cause their neighbors to collapse, and the avalanche proceeds until there are no more newly-collapsed countries (Fig. 1D).

Key quantities in the resulting dynamics are the number of subsequently collapsed countries starting from a given country's collapse (hereafter we call this avalanche size $A$ of the country), and the distribution of avalanche sizes $P(A)$ over all starting nodes [31]. The avalanche size can be used to assess an individual country's potential impact as the crisis epicenter. The dynamics of the model depends on the parameter ratio $f/t$. Depending on the value of the reduced parameter $f/t$, the avalanche size distribution $P(A)$ of all countries takes a qualitatively different form. When $f/t$ is too small, there is no large avalanche, so $P(A)$ decays rapidly with $A$. In contrast, when $f/t$ is too large, there is an excess of global avalanches giving rise to a finite peak around the network-spanning avalanche sizes. In between, there exists a critical point, estimated to be $f/t \approx 7$, at which $P(A)$ becomes power-law-like and the system exhibits the broadest spectrum of avalanche outcomes (Fig. 2). Due to lack of comprehensive records, there is currently no clear empirical evidence of the power law in the magnitudes of economic crises. However, many social and economic phenomena are known to exhibit such a power-law pattern, such as the size of wars measured by the number of casualties [32] and the magnitude of short-term returns of stock market indices [33], as well as some natural and technological phenomena such as the magnitude of earthquakes [31] and the volume of Internet traffic fluctuations [34]. From a methodological viewpoint, the dynamics at the critical parameter provides additional advantage that the avalanche sizes of countries disperse the most, thereby facilitating the clearest comparison between countries. Therefore we focused this study on avalanche dynamics at the critical point.

**Results**

The GDP of a country cannot fully account for its avalanche size

One would expect that a country's avalanche size would relate to a macroeconomic index such as the gross domestic product (GDP) [35]. Indeed, the avalanche size roughly scales with GDP, especially for large GDP countries (Fig. 3). However, the GDP cannot fully explain avalanche size. Specifically, the correlation between them is significant but imperfect: the Spearman rank correlation coefficient is 0.51 ($P$-value = $1.07 \times 10^{-13}$, $t$-test). There are outliers deviating from the overall trend. For example, Spain has a small avalanche size ($A=0$) compared with its GDP (997 billion US dollars) whereas the Republic of South Africa (RSA) has a much larger avalanche size ($A=8$) for its GDP (199 billion US dollars).

To facilitate an understanding of the underlying mechanism for these trends and deviations, we introduce the trade volume-GDP profile (TGP) for a country, which is a scatter plot of trade volume to each trading partner of the country with respect to the partner's GDP (Figs. 4 and S2). For example, the difference between the TGP of Spain and that of RSA is obvious (Figs. 4A and B). Spain is a good example of what we call a proportional trader, engaging in more trades with high GDP countries and less trades with low GDP countries. On the other hand, the RSA concentrates its trade on low-GDP countries that have low capacities, and this brings about the collapse of countries located far above the diagonal of the TGP relation (black squares in Fig. 4B). From TGP analysis, it can easily be seen that fluctuations in proportional trading in the low GDP regions signify the degree of potential direct avalanche impact of a country.

Cascade through weak channels entangles the crisis spreading

A crisis spreads in cascades, not necessarily limited to direct impact. For some countries such as France and Hong Kong (Figs. 4C and D), collapsed countries are not only located above the diagonal of the TGP but also in the middle of it, implying the existence of the cascade process. Consider a cascade process with the avalanche starting from Hong Kong as an illustrative example (Fig. 5). The cascade process involves two operationally different mechanisms. It can proceed directly by following successive direct trade channels (via the solid arrows in Fig. 5), or indirectly by following detours (via the dashed arrows in Fig. 5). An indirect cascade propagates through weak links, the weight of which is insufficient to transmit the cascade directly. However, when the impact through such a weak channel combines with impacts through detours, the aggregate impact can be strong enough to transmit the cascade. The weak links have been shown to play an important structural role in financial market network [36]. The fundamental unit of such an indirect cascade is a triad relation with one strong link and two weak links (highlighted with black arrows in Fig. 5). These indirect cascades render the avalanche process nontrivial. Such entangled domino effects can also establish invisible links, through which a crisis can spread even to countries that are not directly connected to the initiation node. For example, a crisis starting from Hong Kong can reach Myanmar, which does not engage in a direct trade relation with Hong Kong (Fig. 5; see Fig. S3 and Table S2 for more examples). In this way, a country's impact on crisis spreading is greatly amplified by the local clustering of weak links, as exemplified by the countries in the Southeast Asian block, similarly to the accelerated spread of behavioral adoption through clustered social networks [37].

Such a nontrivial cascade effect accounts for a significant fraction of the avalanche for countries with relatively large avalanche sizes. To assess this effect quantitatively, we divided a full avalanche process into four sub-processes (Fig. 6): i) A one-step direct avalanche comprised of countries that collapse by direct impact from the starting country; ii) a multi-step direct avalanche comprised of countries that collapse by direct cascade processes; iii) an indirect avalanche comprised of countries that collapse through indirect cascades; and iv) a residual avalanche comprised of countries that collapse through impacts from countries in indirect avalanche. For most countries with an avalanche size $A > 10$, the indirect avalanche constitutes more than half of the total avalanche (Fig. 6). This clearly shows that crisis spreading might not always propagate linearly and could occur in a highly entangled way. Therefore, it is crucial to take the entire network into account in order to fully understand the generic feature of the crisis spreading process. Country-specific local connectivity patterns can also affect an individual country's avalanche profile. For example, Russia shows a higher fraction of one-step direct avalanches, most of which are former members of the Soviet Union with low economic capacities. Hong Kong, Indonesia, and Malaysia exhibit higher fractions of multi-step direct avalanches that originate from their strong connectivity to Singapore.

The duration of an avalanche reveals additional aspects of the entangled propagation of crisis spreading. The avalanche duration of a country is defined as the number of cascade steps needed for an avalanche process starting from the country to come to an end. In general, larger avalanches tend to occur over longer durations, and smaller ones tend to be shorter (Fig. 7). However, significant deviations from this simple trend could occur. For example, the RSA has a very short duration with respect to its avalanche size. Its impact is entirely confined to its immediate South African neighborhood and the avalanche proceeds for only two steps. In contrast, countries such as Hong Kong, Malaysia, and Singapore have much longer durations of six steps compared to other countries with similar avalanche sizes. This behavior is again rooted in strong weak-channel clustering prevalent among these Southeast Asian countries, which can support complicated pathways with alternating direct and indirect cascades.

Large-scale patterns of crisis spreading

Characterizing the overall structural properties of the global economic system is necessary to reveal the global-scale relation between the network structure and the dynamics of crisis spreading. To this end, we built a spanning tree for the GMN comprised of maximum weight links [38]. As shown in the spanning tree representation (Fig. 8A and Text S1), the global economic system exhibits continental clustering, which has widely been acknowledged in the literature [24, 39]. In addition, we built what we call the avalanche network, in which two countries are connected by a directed link if a country can make the other country collapse, resulting in 367 directed links connecting 175 countries (Fig. 8B; see Fig. S4 for a more detailed map with annotations). In the avalanche network, 12 countries, such as Bhutan and Solomon Islands, are isolated. These countries are effectively economically isolated and engage in very limited trade with the outside, so they neither cause other countries to collapse nor collapse due to other countries. As expected, the overall aspect of avalanche relations dictated by the avalanche network followed the continental clustering structure of the GMN. To assess the degree of continental clustering at the GMN and the avalanche network structures, we calculated the fraction of links connecting countries within the same continent (colored links in Fig. 8). In the spanning tree of the GMN, 97 out of 174 links (56%) are intra-continental links. In the avalanche network, 150 out of 367 links (41%) are intra-continental links. As we will show, these fractions are significantly high, quantitatively confirming the presence of continental clustering. This continental clustering property enabled us to build a coarse-grained avalanche network at the continental level, revealing the major flow of influence and dominance over different continental regions (Fig. S5). We note that the large-scale pattern in the continental avalanche network resembles the map of information flow based on the causal relations between stock market indices described in [40], indicating that both product exchange and information flow between markets commonly reflect their economic interdependency.

Comparison with randomized network structures

To further investigate the effect of the network structure of the global economic system on crisis spreading dynamics, we introduced two randomized networks and studied their crisis spreading characteristics against that of the GMN. First, we built a globally-shuffled network (GSN) in the following way: i) Choose two links randomly. ii) Consider swapping the trade partners of the two links. iii) If the changes of both new links' weights from the original links' weights in the GMN are less than 1%, accept the exchange. Otherwise the exchange is not accepted and restored. iv) Repeat i)-iii) until the stationary state was reached (see Text S1 and Fig. S6 for more details). An important point is to refer the original link weight rather than the current weight for testing the exchange acceptance criterion to prevent the accumulated changes of link weights that may distort a country's trading profile undesirably. Using this randomization approach, we were able to shuffle each country's trading partners randomly while maintaining the number of links and trade volumes of each link of each country. This randomization has the effect of untangling the continental clustering. Indeed, the spanning tree of the GSN (Fig. 8C) typically contained only 35±3 (mean ± standard deviation) intra-continental links (20%), a significant reduction in continental clustering by more than one-half (empirical $P$-value < $10^{-3}$). The reduction of continental clustering was also found in the avalanche network over the GSN (Fig. 8D), in which only 80±3 out of 608±18 links (13%) were intra-continental links; again, a highly significant decrease by more than one-half compared to the GMN (empirical $P$-value < $10^{-3}$).

The untangling of continental clustering in the GSN had the significant effect of increasing the overall degree of crisis spreading. In the GSN, the sum of avalanche sizes of all countries was 608±18, significantly larger than the sum 356 in the GMN (empirical $P$-value < $10^{-3}$). The origin of this enhanced crisis spreading is that due to the untangling of continental clustering, the failure of major countries could spread farther and more broadly than in the GMN. Reflecting this, developed countries such as the United States, Germany, Japan, China, and France exhibited considerably larger avalanche sizes in the GSN than in the GMN, which dominantly accounted for the increased sum of overall avalanche sizes (Figs. 9A, 9B, and 10A). On the other hand, many countries had their avalanche sizes reduced in the GSN, such as Hong Kong and Malaysia, and some countries completely lost their spreading ability, such as the RSA (Fig. 10A). Most of avalanche size changes, both the increase and the decrease, are accounted for by the indirect avalanche (Fig. S7). This means that the nontrivial effect of the local connectivity structure was almost disentangled in the randomized networks, and the avalanche dynamics became simplified for most countries. Only when the gross impact of a country exceeded a threshold did crisis spreading through indirect avalanche occur and dominate. In this sense, the rich-get-richer and the poor-get-poorer aspects became more severe in the GSN.

Polarization of impact was further intensified in the second version of our randomized network−the so-called globally-distributed network (GDN). In the GDN, each country can have as many trading partners as is constrained only by the total trade volume. Specifically, we constructed a GDN in the following way: i) Discretize and divide each trade link of GMN into unit links in one million US dollars. ii) Start from the network with all the unit in- and out-links unlinked. iii) Select one export unit and one import unit from the unlinked unit links. iv) If the export and the import countries are different, connect the two by an arrow. Otherwise, discard the trial. v) Repeat iii)-iv) until all unit links are connected. v) Merge unit links with same export and import countries, restoring the weighted network structure. In the hypothetical globalized world represented by a GDN, the avalanche size distribution becomes even more polarized. In the GDN, only three countries−United States, Germany, and China−have dominant avalanche sizes (Fig. 9C). Yet, the average avalanche size of these three countries is 132, spanning as much as 75% of the globe (Fig. 10B).

To quantify the degree of polarization in the avalanche impact for different network structures, we calculated two quantities: the typical size, and the likelihood of nonzero avalanches. The former is given by the average of avalanche sizes over countries with nonzero avalanche sizes and describes the expected global impact that the global economic system might suffer once the avalanche occurs. The latter is given by the number of countries with nonzero avalanche sizes divided by the total number of countries, providing a measure of how likely an avalanche is to occur when crises are initiated randomly. Both randomized networks have a higher typical size, $20\pm1.8$ for the GSN (empirical $P$-value $< 10^{-3}$) and $132\pm11.0$ for the GDN (empirical $P$-value $< 10^{-3}$), compared to 9.7 for the GMN (Fig. 10B). The finding that spreading of crisis becomes larger for random networks is consistent with the results from a recent modeling study based on epidemic spreading [41]. On the other hand, they have a lower likelihood, 0.18±0.015 for the GSN (empirical *P*-value = 0.007) and 0.017±0.002 for GDN (empirical *P*-value < $10^{-3}$), compared to 0.22 for the GMN. This means that crisis spreading might take place less often in randomized economic systems, but once it occurs, its impact would be far more extensive and widespread than it is now in the GMN. Underlying this robust-yet-fragile property is the dual role of big countries: On one hand, they act as shock absorbers by tolerating and buffering impacts from most countries' collapse with their large reservoir of economic capacities and alternative connection channels [28]. At the same time, they are also the Achilles' heel [42] of the global economic system in the sense that their failure would trigger the spreading of a worldwide crisis, leading to an almost complete failure of the global economic system. From the network topological perspective, the randomized structures (the GSN and the GDN) possess elevated degrees of a rich-club effect [43] compared to the GMN through the disentanglement of continental clustering (Fig. S8). Such enhancement of mutual linkage between large economies in the randomized structures therefore may in part be responsible for the changes in the crisis spreading property. This double-faceted picture emerging from the crisis spreading dynamics of the global economic system can be a starting point from which more reasonable international relations development scenario based on cascade control [44] and optimization for network resilience [45] under crisis spreading can be formulated.

Robustness of the model results

The specific outcomes of avalanche dynamics may well depend on the fine-scale structure of network linkages and the mechanistic details of the crisis spreading process, modulated by episode-specific environmental factors. Our toy model of crisis spreading may have neglected many potentially important factors in crisis spreading dynamics. Therefore, it is important to confirm the robustness of the general features under economically reasonable modification of the model's details. Here we will present one such modification of the model. Results of other modifications are deferred to Supporting Information. Tolerance to crisis of a country may be dependent also on the fiscal conditions of the country that are not directly reflected in the GDP. For example the current account balance (CAB) of the country may be a simple indicator of its macroeconomic fiscal soundness. We incorporated CAB to the model dynamics for investigating the robustness of the main conclusions. To this end, we modified the economic capacity of each country to be the sum of its GDP and CAB, such that countries with positive (negative) CAB will tolerate the impact from crisis to a larger (lesser) extent. We obtained the CAB data from the International Monetary Fund World Economic Outlook Database, from which we also obtained the GDP data. Under these modification, the resulting overall feature of the model dynamics is similar to that of the original model (Fig. 11): the power-law-like avalanche size distribution is obtained around $f/t \approx 7$ (Fig. 11A); the avalanche profile exhibits large portions of indirect avalanches (Fig. 11B); and there are significant intra-continental linkages, that is 146 out of 519 links (28%) in the avalanche networks (Fig. 11C).

We have tried a number of other modifications, including link and node weight rescaling, different cascading rules, and taking other country-specific fiscal conditions into account (Text S1, Figs. S9 and S10). Under all these modifications, we confirmed that the overall patterns of the crisis spreading dynamics described in this study remained unaltered. Furthermore, these overall patterns were robustly and consistently obtained for a wide range of the model parameter $3 < f/t < 20$ (Text S1 and Figs. S11-S14). Therefore, the global pattern of crisis spreading dynamics obtained in the study is likely robust and generic. There are still missing components in our modeling; additions of other economic channels, such as the financial links which are thought to be a major origin of the current turmoil, would result in a better episode-specfic explanation of the sequence of crisis spreading dynamics. They are, however, not expected to alter the overall avalanche properties qualitatively either, as long as they do not completely change global network characteristics such as regional and continental clustering. Therefore, even though the global economic system will inevitably undergo structural and environmental changes over time, we expect that the general framework and conclusions established in this study will be relevant to the systemic investigations of global economic problems.

## Discussion

We have studied the crisis spreading dynamics on global economic system using a simple toy model of crisis spreading on top of the global macroeconomic network built from the international trade data. Focusing on the role of the network topology at the local and global level, we have shown that the impact of a country to the spread of crisis is not fully captured by its simple macroeconomic index such as GDP, but its connectivity profile is also instrumental for a better understanding. Beyond the direct impact, the indirect impacts propagating through weak links form a significant part of the avalanche process of crisis spreading. At the local or regional scale, we have shown that the strong regional blocs leading to clustering of weak links can aggravate the crisis spreading, by accumulating impact through the dense multilateral connectivity within the blocs. At the same time, on the global scale, the current structure of global macroeconomic network harbors higher tolerance to extreme crises than the more "globalized" network structures obtained by randomized global networks. These results may have an interesting implication on the hidden cost of the ongoing globalization movement: In a more globalized macroeconomic network in which the regional and continental clustering continues to become untangled via establishing free trade agreements transcending regional blocs, a crisis would propagate throughout the world economy more easily, and therefore the global economy would become more susceptible to extreme crises and potentially an unprecedented level of world-wide depression.

It is now widely accepted that a systems or a network view is essential for a better understanding of global economic challenges [46, 47]. The increasing availability of various data at the global scale, and tools and insights from network science with which to analyze them, provide an opportunity to transform this conceptual necessity into a quantitative possibility. The network dynamics approach undertaken here has demonstrated utility and promise, offering a general picture of crisis spreading dynamics at the global network level and at the individual country level. We hope that our simple model and analysis trigger the interest of researchers from diverse disciplines, including international economics and finance, systems and ecological engineering, and the physical and mathematical sciences, thereby helping them cooperate to better understand and improve the global economic system.

## Acknowledgments

We thank Shlomo Havlin, Cesar Hidalgo, Jeho Lee, and Didier Sornette for their helpful comments.

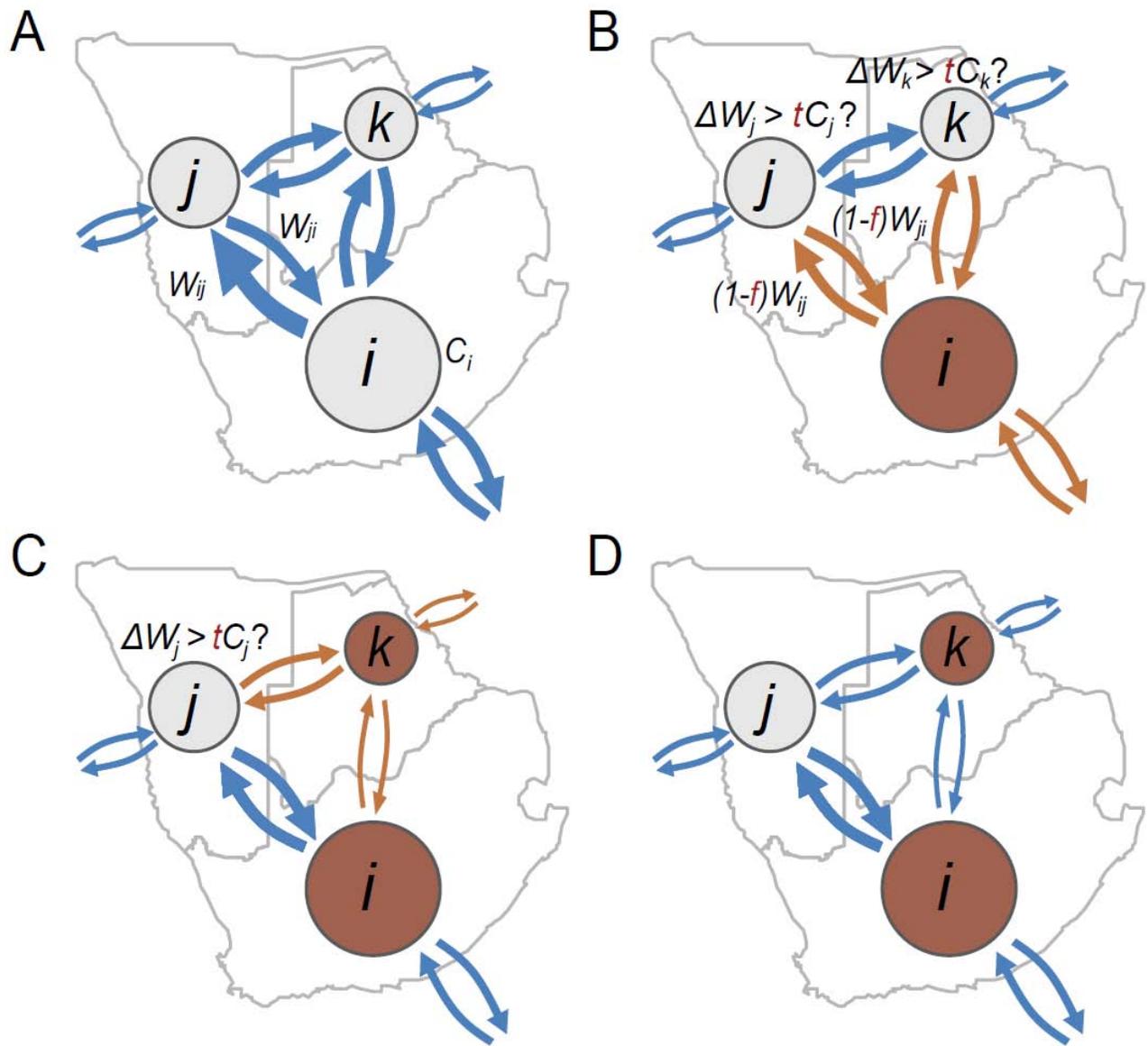

**Figure 1. Crisis spreading model.** The dynamics of the crisis spreading model proceeds as follows: (A) Each node (country) is assigned a capacity $C$ given by its GDP value, and two countries $i$ and $j$ are connected by a directed link given by the trade relation with the associated weight $W_{ij}$ given by the corresponding trade volume. (B) Suppose that a country $i$ collapses (indicated as dark brown). Then the weights of all links of country $i$ (indicated as brown) are decreased by a fraction of $f$. And the total decrement $\Delta W$ of either the incoming or outgoing link weights of any countries connected to the collapsed country is compared with the fraction $t$ of its node capacity. (C) If the total decrement exceeds the threshold for any countries, these countries also collapse (country $k$, in this example), triggering an avalanche of collapses. The link weights of newly-collapsed countries are decreased by the fraction $f$, and total decrement $\Delta W$ of each country is re-evaluated. (D) This process continues until there are no more newly-collapsed countries.

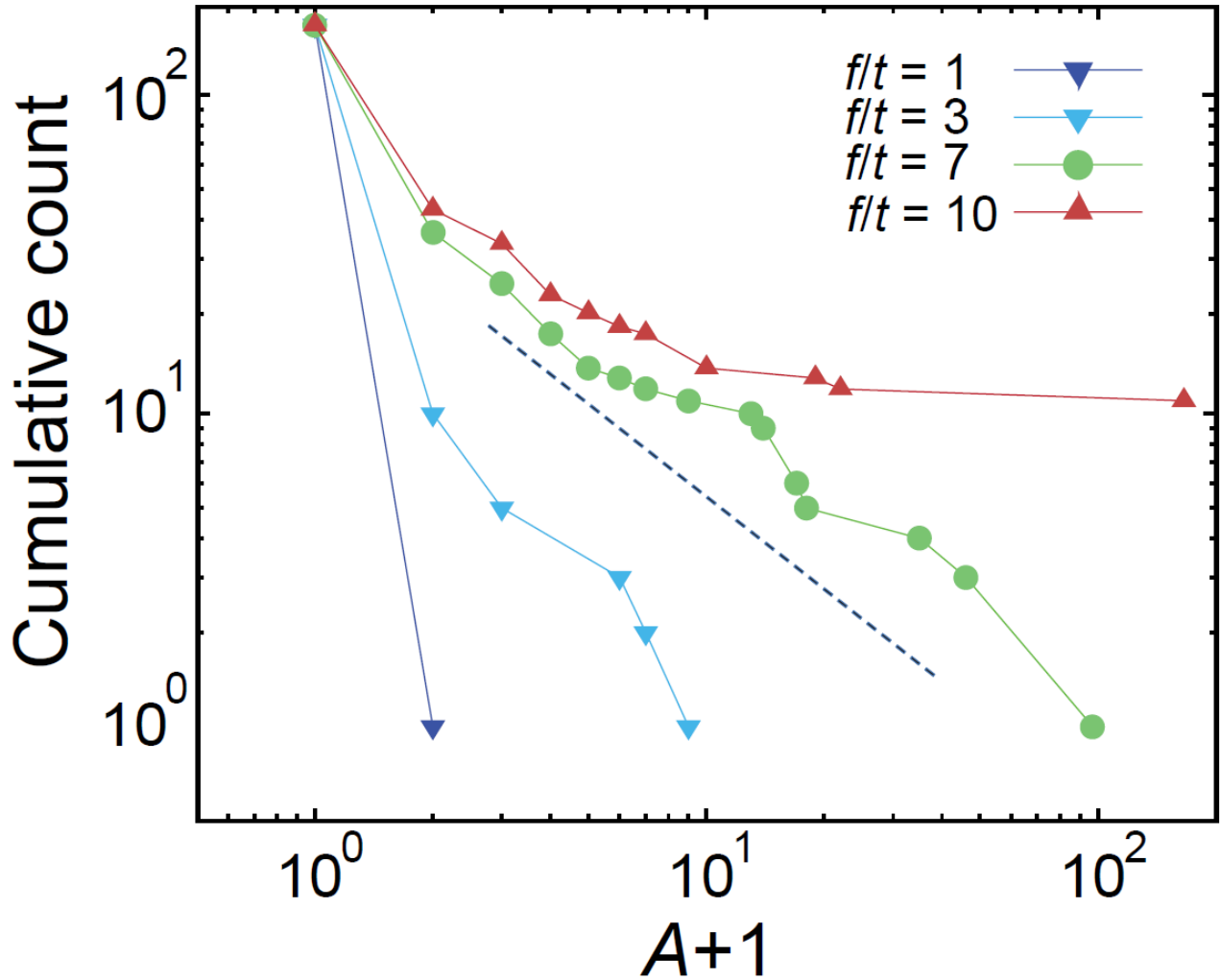

**Figure 2. Avalanche size distributions.** The cumulative counts of countries with avalanche sizes equal to or greater than *A*, which behaves in the same way as the cumulative distribution of *P(A)*, is plotted. At the critical parameter *f/t*=7, the plot becomes straight with a slope of -1 (dashed line) on a logarithmic scale, indicating the power-law relation $P(A) \sim A^{-2}$. The horizontal axis has been offset by one, to plot the countries with *A*=0 on a logarithmic scale.

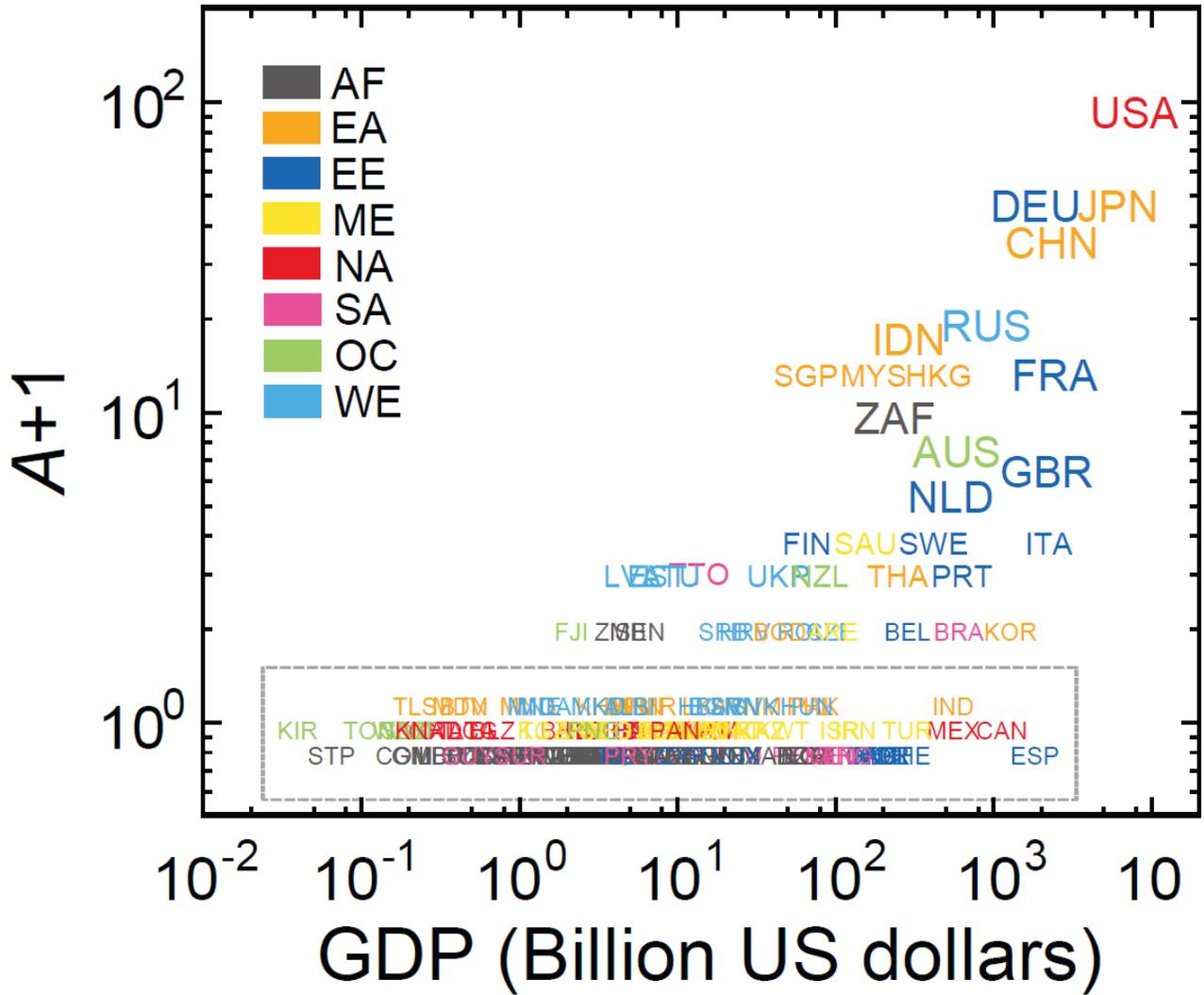

**Figure 3. Avalanche size vs. GDP.** Scatter plot of the avalanche size of each country offset by one ($A+1$) as a function of its GDP is displayed with the color code by continental association (AF: Africa, EA: East Asia, EE: East Europe, ME: Middle-East, NA: North America, SA: South America, OC: Oceania, and WE: West Europe). The overall increasing trend with some deviations indicates that the avalanche size of a country is partly accounted for by the GDP, but not entirely. The dashed box indicates a group of countries with $A=0$ (split to minimize overlaps to enhance visual comprehensibility). Country name codes follow the ISO 3166-1 alpha-3 code throughout the paper.

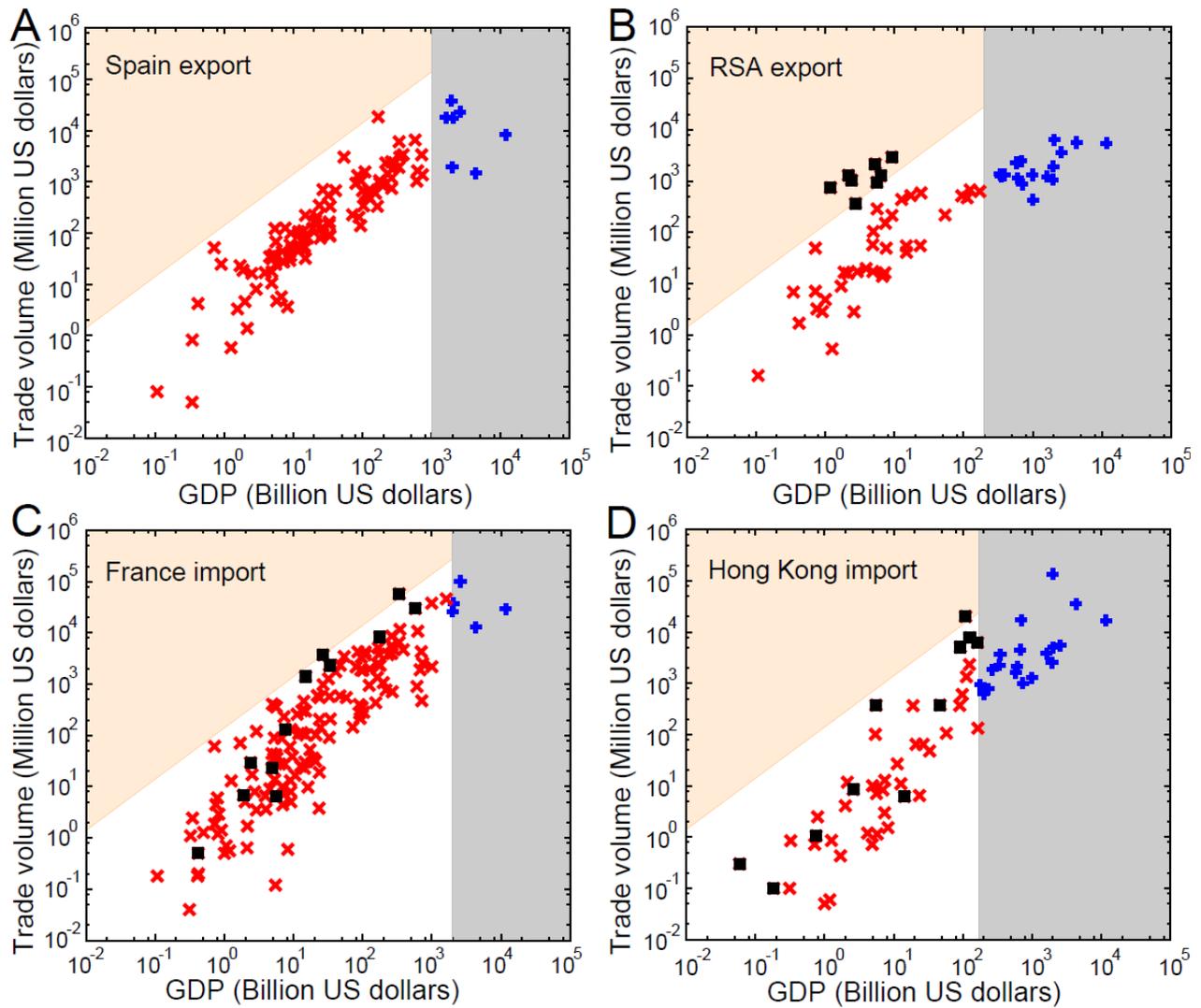

**Figure 4**. **Trade volume-GDP (TGP) profile.** TGP of (A) Spain, (B) the RSA, (C) France, and (D) Hong Kong are displayed. Red points denote the countries with lower GDPs than the starting country, and blue points in the gray area are those with higher GDPs. The black squares represent countries that are collapsed by the starting country. The area shaded in orange denotes the region within which countries can be affected directly by the starting country.

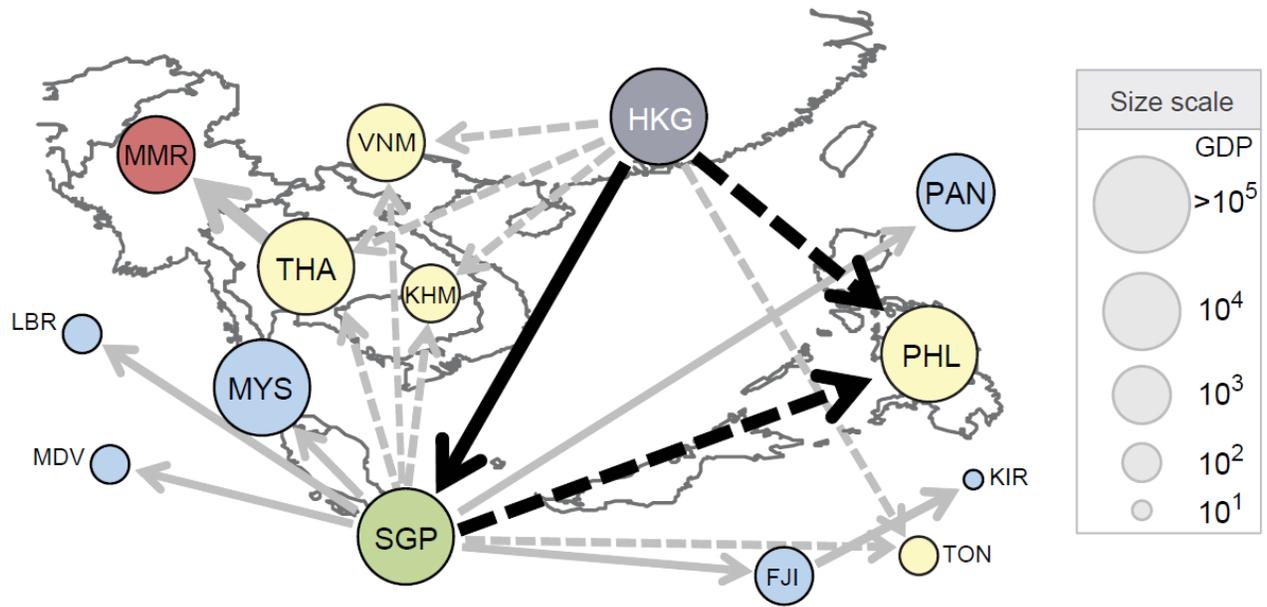

**Figure 5. Full sequence of avalanche process starting from Hong Kong.** Direct channels (solid arrows) and indirect channels (dashed arrows) are distinguished because they contribute to the avalanche process by different mechanisms. Countries are colored according to the sub-process they belong to, and their size is given by the GDP (in million US dollars). The starting country, Hong Kong, is colored in gray.

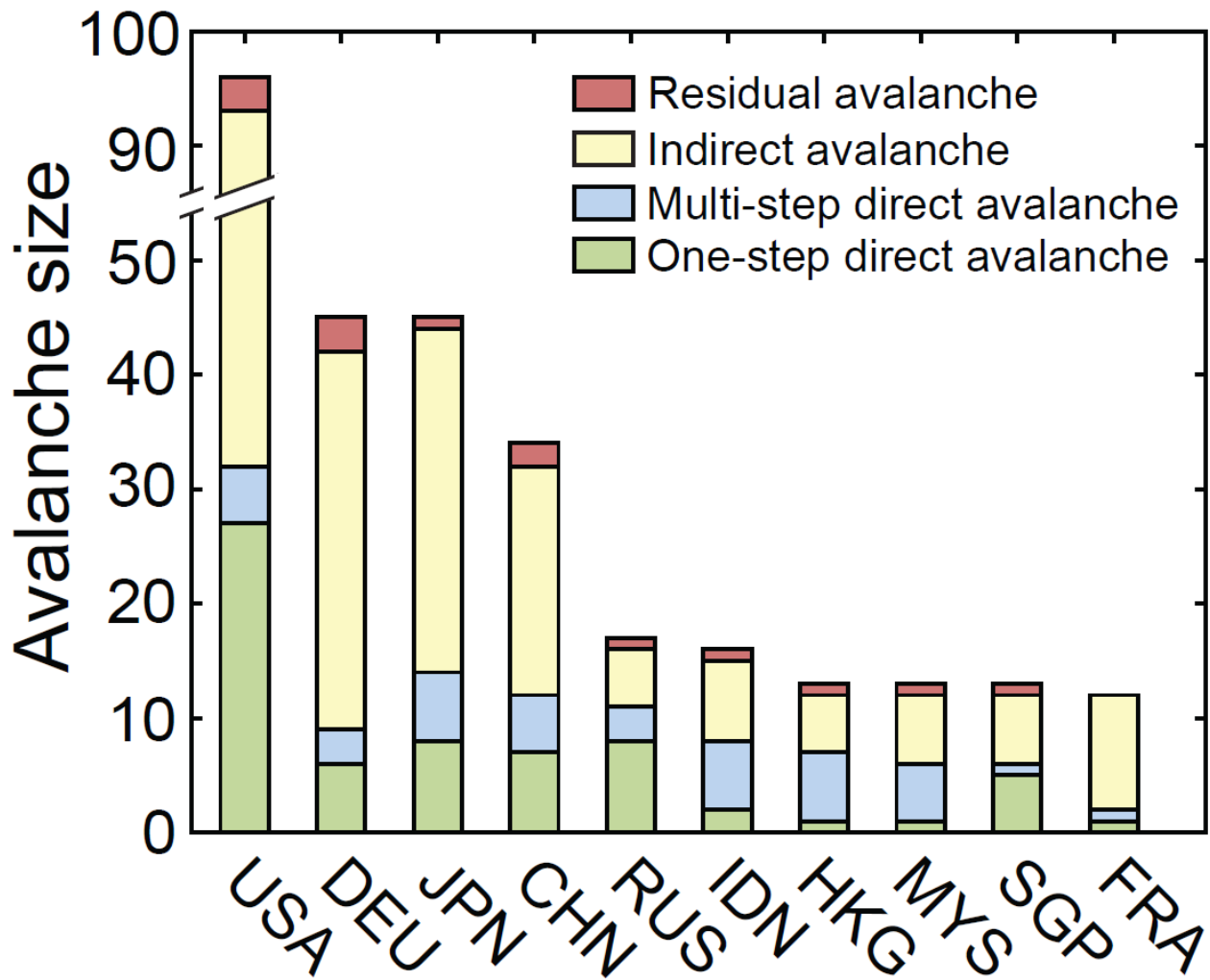

**Figure 6. Avalanche profiles.** Bar plot showing the avalanche profile of countries with the ten largest avalanche sizes is displayed. The total avalanche process is divided into four sub-processes and the colored bar denotes their distribution. For most countries shown in the figure, the indirect avalanche (yellow) constitutes the largest fraction of the total avalanche process.

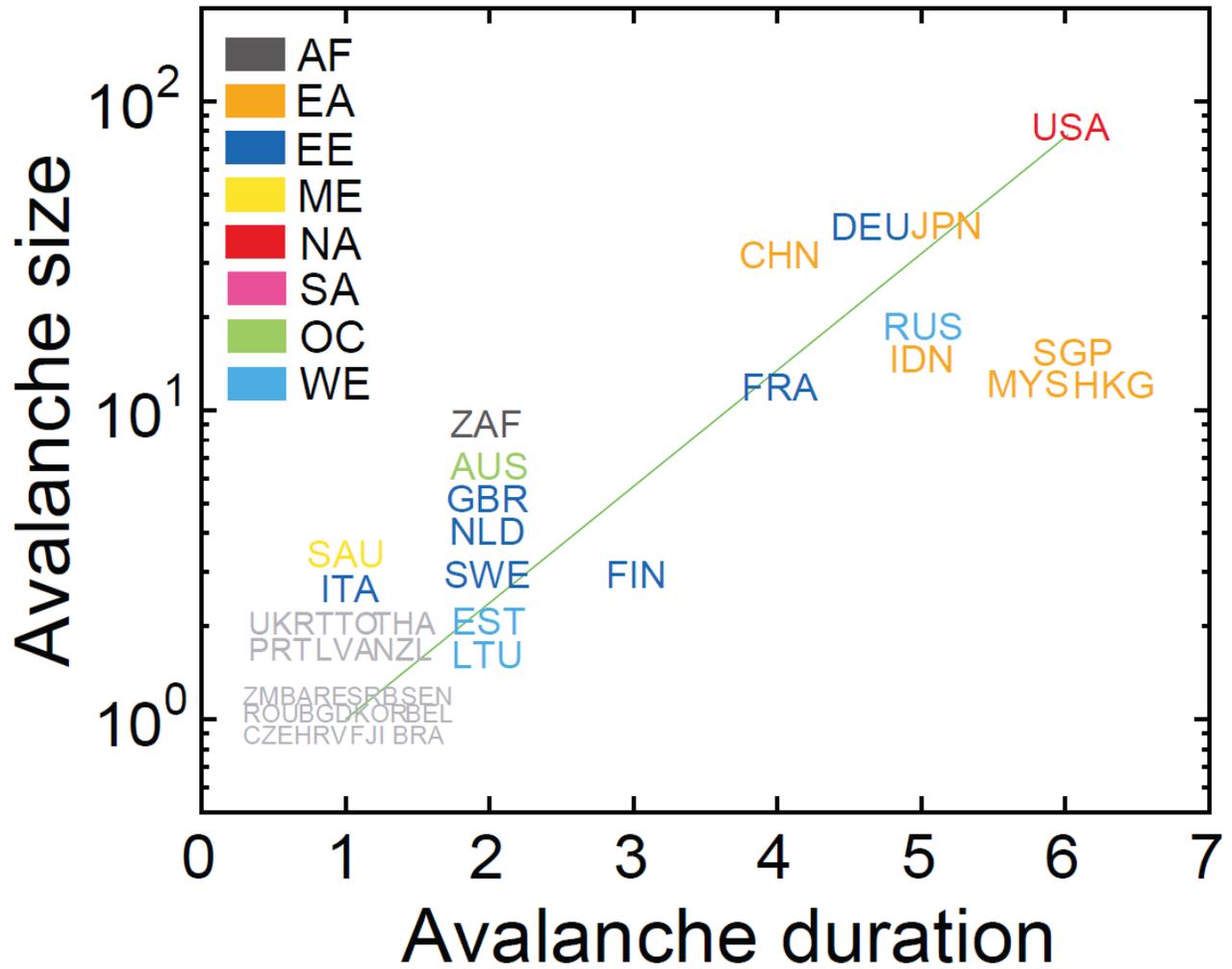

**Figure 7. Avalanche durations.** Relation between avalanche duration and avalanche size for 175 countries is displayed. Note that some countries have much longer or shorter durations compared to their avalanche sizes, thereby deviating from the overall increasing trend. The same color code for the continental associations as in Fig. 3 are used.

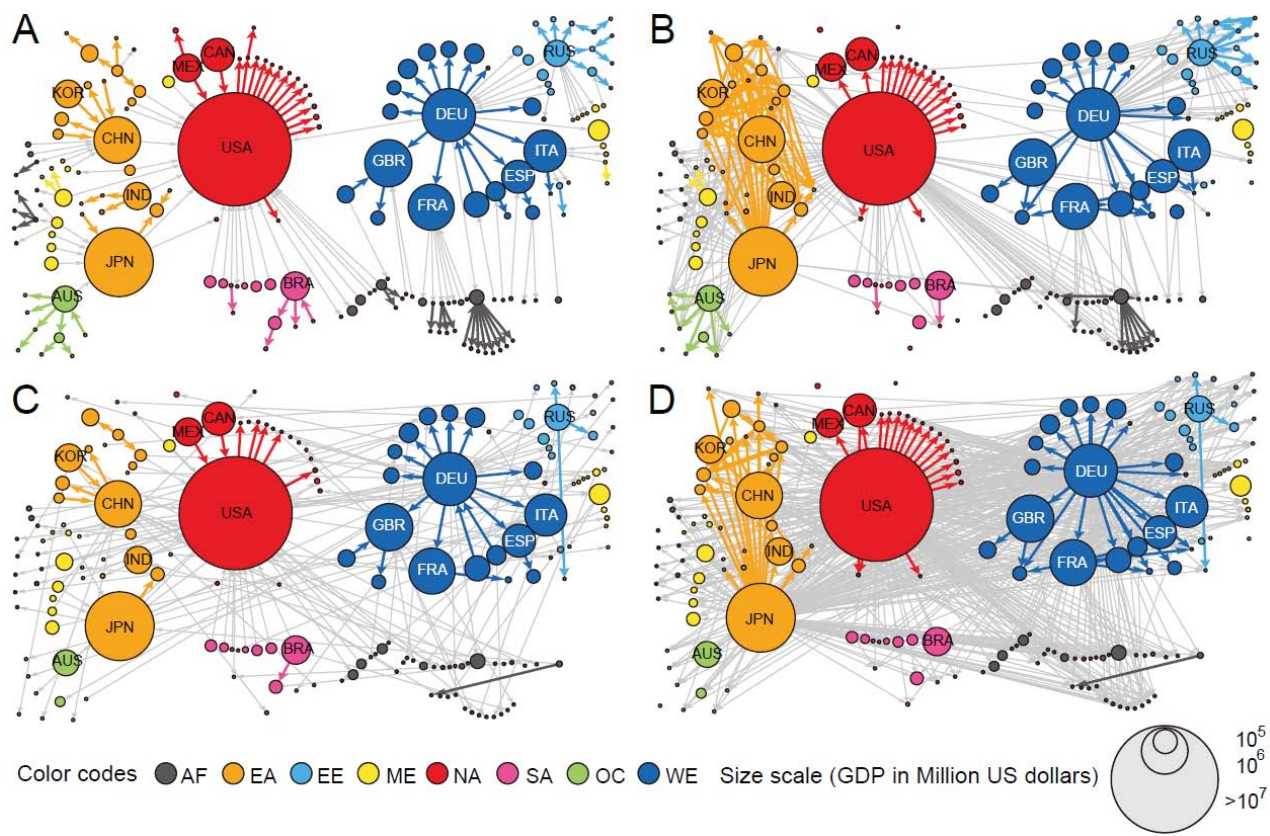

**Figure 8. Global patterns of crisis spreading.** (A) Spanning tree representation and (B) the avalanche network of the GMN are displayed together with (C) the spanning tree and (D) the avalanche network of a typical sample of the GSN. The intra-continental links are colored according to the color code of the continent and inter-continental links are colored in gray. A more detailed version of the avalanche network of (B) is available as Fig. S4.

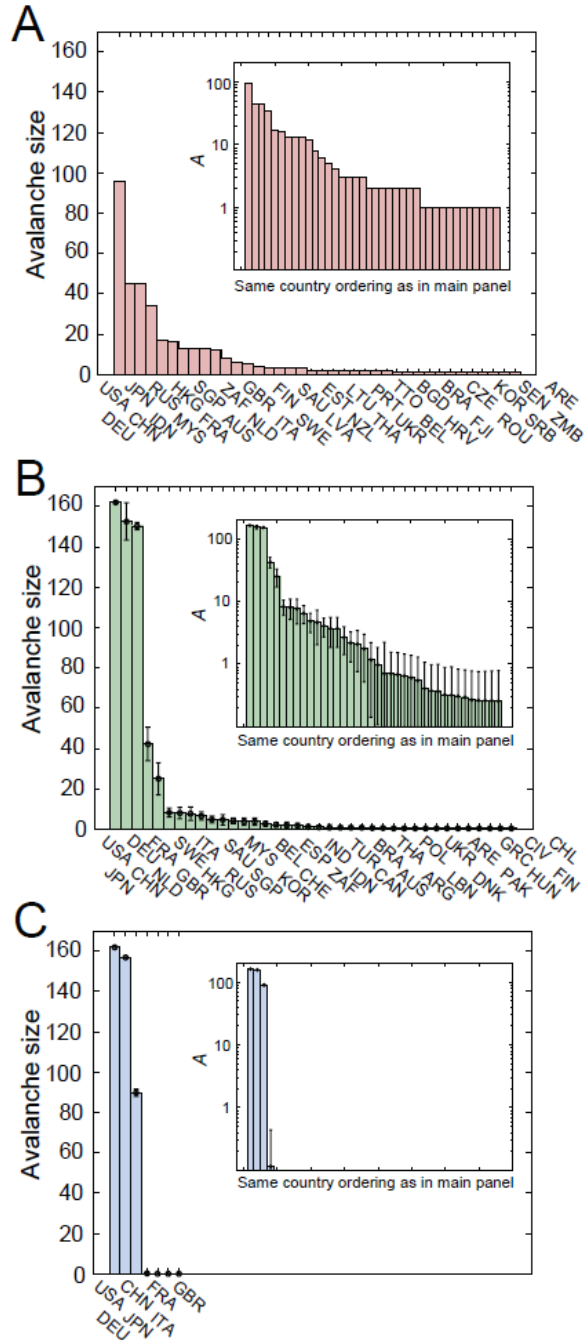

**Figure 9. Comparison of avalanche sizes in different network structures.** Bar plots of avalanche sizes in decreasing order for (A) GMN, (B) GSN, and (C) GDN are displayed. (Inset) The same plots in semi-logarithmic scales are displayed. Results for both the GSN and GDN were obtained from $10^3$ different randomized networks. Error bars represent the standard deviations over these samples.

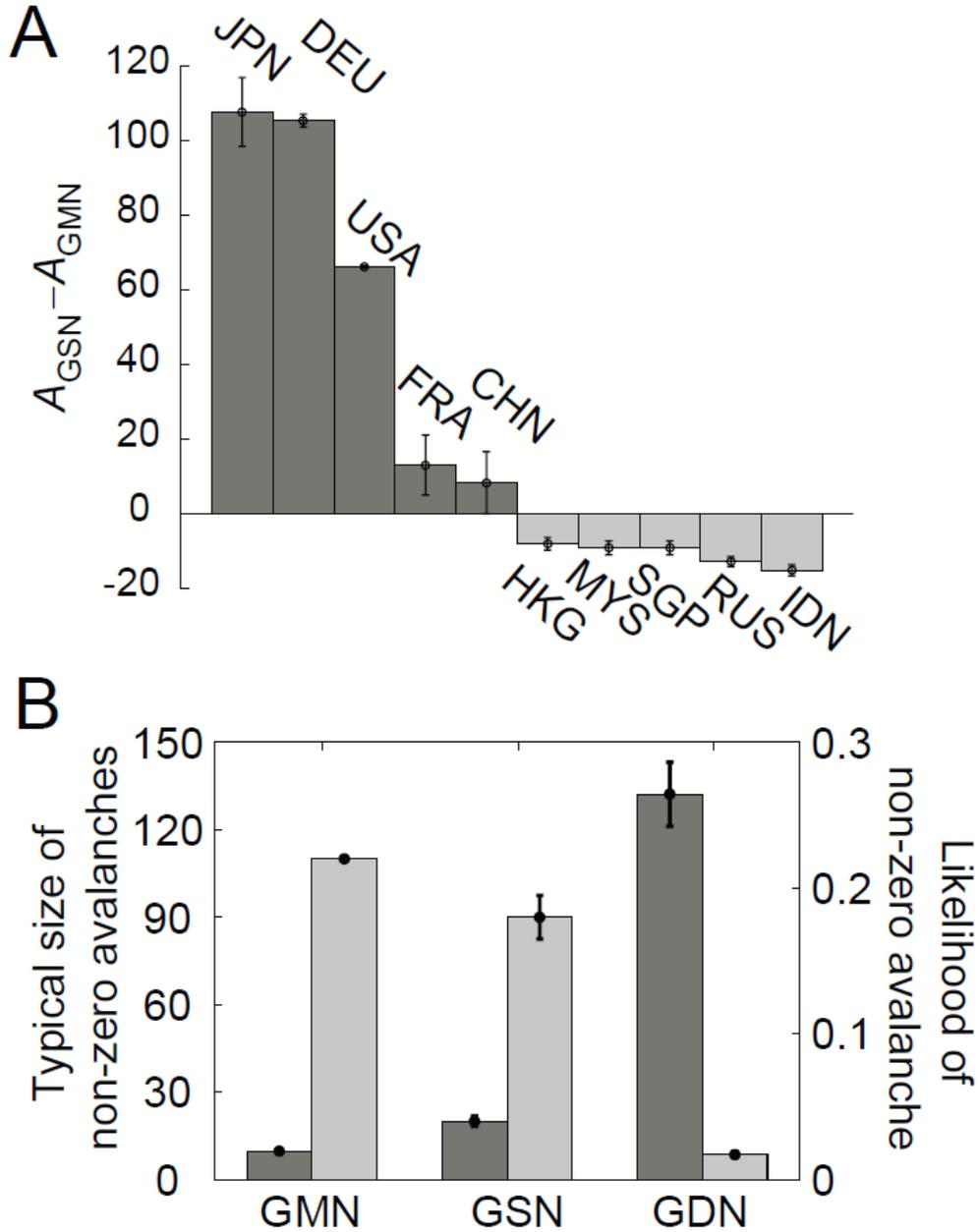

**Figure 10. Comparison of crisis spreading in different network structures.** (A) Changes in the avalanche size for each country in the GSN with respect to that of the GMN. The avalanche size increased for the United States, Germany, Japan, China, and France, and decreased for other countries. (B) Comparison of typical nonzero avalanche sizes (dark gray) and the likelihoods of an avalanche (light gray) for the GMN, GSN, and GDN. Results for both the GSN and GDN were obtained from $10^3$ different randomized networks. Error bars represent the standard deviations over these samples.

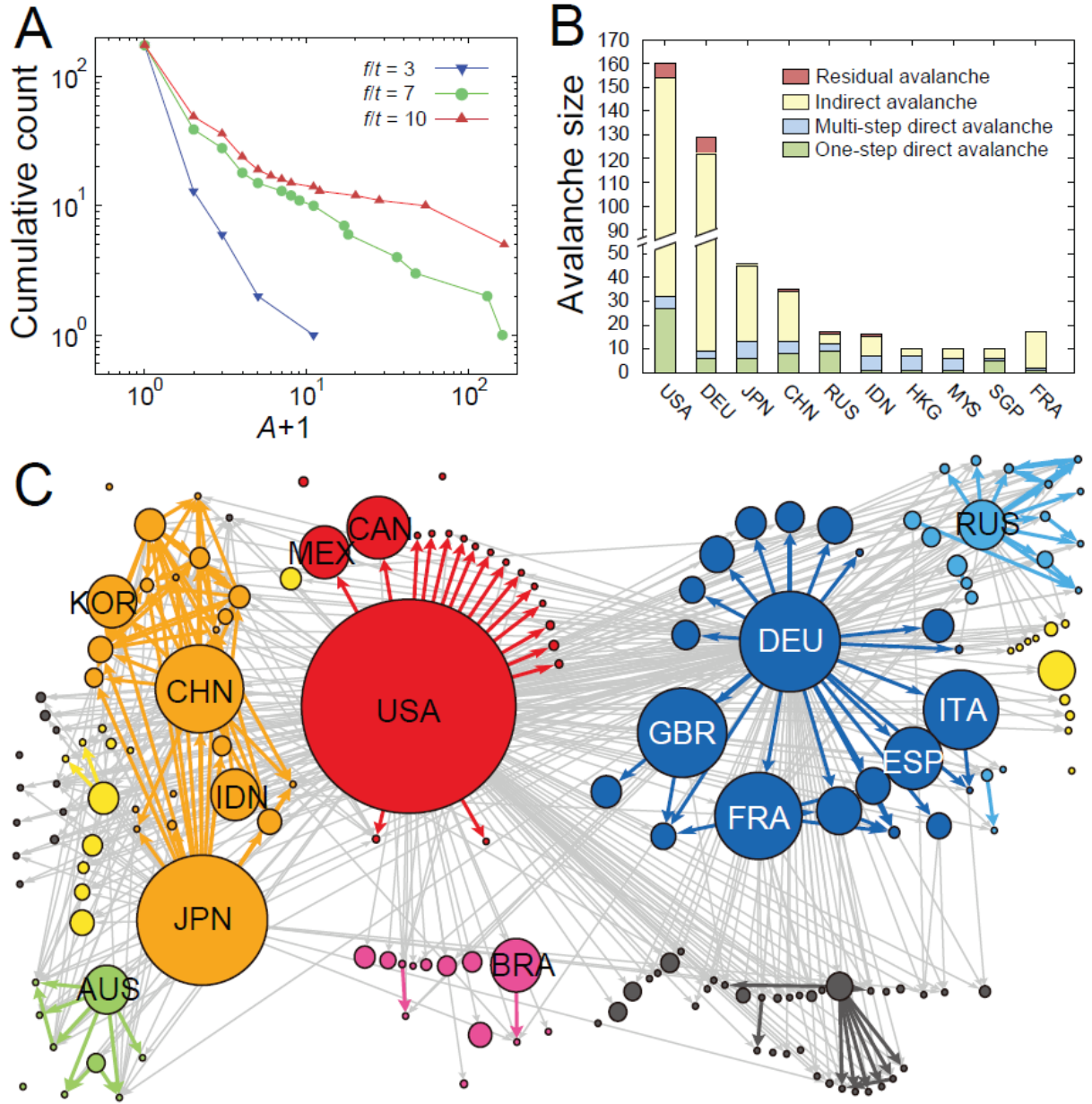

**Figure 11. Results of the modified crisis spreading model with CAB.** Displayed are (A) the cumulative counts of avalanche sizes, (B) the avalanche profile of countries with the ten largest avalanche sizes, and (C) the avalanche network of the modified model. One may note that similarity of the overall results to those of the original model (Figs. 2, 6, and 8B, respectively), despite some quantitative changes in the numeric values. In the modified model, we recover the power-law-like $P(A)$ at $f/t \approx 7$; the indirect avalanches constitute the dominant part of the avalanche profiles; and the avalanche network maintains the continental clustering pattern.